\documentclass[runningheads]{llncs}
\usepackage{amsmath}
\usepackage{graphicx}
\usepackage{float}
\usepackage[utf8]{inputenc}
\usepackage{fourier}
\usepackage{array}
\usepackage{makecell}
\usepackage{url}
\usepackage{algorithmicx}
\usepackage{algorithm}
\usepackage[noend]{algpseudocode}
\usepackage{amsmath}
\usepackage{color}
\graphicspath{ {images/} }
\usepackage{xcolor}
\usepackage{pgfplots}
\usepackage{listings}
\usepackage{multirow}
\usepackage{tikz}
\usepackage{subfig}
\usepackage{romannum}
\usepackage{pgf}
\usepackage[utf8]{inputenc}
\usepackage[strings]{underscore}
\usepackage{url}
\usetikzlibrary{arrows,automata}
\usetikzlibrary{positioning}
\usepgfplotslibrary{statistics}
\newcommand*{\rom}[1]{\expandafter\@slowromancap\romannumeral #1@}
\usepackage[title]{appendix}
\usepackage[final]{pdfpages}
\algnewcommand{\LineComment}[1]{\Statex \(\triangleright\) #1}

\begin{document}
\title{Distributed Symbolic Execution using Test-Depth Partitioning}

\author{Shikhar Singh \and
Sarfraz Khurshid}

\authorrunning{S. Singh et al.}
\institute{University of Texas, Austin, USA \\
\email{shikhar_singh@utexas.edu}\\	
\email{khurshid@ece.utexas.edu}}
\maketitle              % typeset the header of the contribution
\begin{abstract}
Symbolic execution is a classic technique for systematic bug finding, 
which has seen many applications in recent years but remains hard to scale. 
Recent work introduced ranged symbolic execution to distribute the symbolic 
execution task among different workers with minimal communication overhead 
using test inputs.  However, each worker was restricted to perform only a 
depth-first search. This paper introduces a new approach to ranging, 
called \emph{test-depth partitioning}, that allows the workers to employ different search 
strategies without compromising the completeness of the overall search. 
Experimental results show that the proposed approach provides a more flexible 
ranging solution for distributed symbolic execution.\par
The core idea behind test-depth partitioning is to use a test-depth pair to 
define a region in 
the execution space. Such a pair represents a partial path or a prefix, 
and it obviates the need for complete tests to determine 
boundaries as was the case in the previous ranging scheme. Moreover, different 
workers have the freedom to select the search strategy of choice without 
affecting the functioning of the overall system.  Test-depth partitioning is implemented 
using KLEE, a well known symbolic execution tool. The preliminary results 
show that the proposed scheme can prove to be an efficient method to speed up symbolic 
execution.

\end{abstract}
\section{Introduction}
Symbolic execution, conceptualized and demonstrated almost 40 years ago\cite{se1,se2}, 
is one of the most versatile and influential methodologies for analyzing software. 
The core of symbolic execution is a technique that undertakes a structured exploration 
of the execution paths which exist in a given program. A standard symbolic execution 
tool comprises of two main components. 
The first component constructs the path conditions, which are constraints on program 
inputs that cause the execution of a particular path.
The second component is a mechanism to solve the path conditions and 
provide concrete values to the program inputs. Symbolic execution has found widespread 
application in test input generation. Solving path conditions for execution paths in a 
program yields a capable test suite which provides better code coverage. Advancements in 
SAT and SMT solving technology coupled with the rapid rise of computing power has paved 
the way for using symbolic execution in a diverse range of real-world scenarios.\par
While symbolic execution presents itself as an attractive tool to test various aspects 
of software design, it suffers from scalability issues. The main reasons for this drawback 
are the complexity of path conditions and the state space explosion as software becomes 
more extensive and more expressive. 
These factors lead to prohibitively high exploration times, which hinders the adoption 
of this technology.
There have been several research endeavors to address these bottlenecks. 
\emph{Concolic} execution, introduced in DART\cite{conc1}, combines concrete and 
symbolic execution to limit the path explosion problem. In this technique, the program 
runs with program inputs having concrete values. The symbolic path conditions for the 
execution path are recorded and the last branch in the path condition is negated to execute 
a new path. Several tools extend concolic execution to target different programming 
environments and provide additional features \cite{conc2,conc3,conc4}. DiSE\cite{dise}, 
using an incremental approach, symbolically executes only the modified parts of the code. 
The compositional approach used in SMART\cite{compo} uses concolic execution for functions 
in isolation to reduce the number of paths. The distributed approach, our focus in this work,
involves breaking 
down a large symbolic execution problem into smaller ones and solving them in a 
parallel/distributed fashion. Cloud9\cite{cloud9,cloud92} is a popular tool which carries 
out distributed symbolic execution and has been applied to test real-world systems. 
Distributed approach has also been used on Java bytecode\cite{jpfpar1,jpfpar2}.\par
Several modern symbolic solvers incorporate the Execution Generated Test\cite{egt1} 
approach where both concrete and symbolic states of a program are maintained. 
If an operation involves all concrete values, then it is executed normally, but in 
the presence of one or more symbolic variables, symbolic execution takes place. 
EXE\cite{test1} and KLEE\cite{klee1} are two well-known tools in this category. 
KLEE is an open source symbolic execution tool that works on LLVM\cite{llvm} 
generated bytecode and integrates with the LLVM architecture. KLEE, and tools 
built on top of KLEE have been used to test a wide range of software applications 
like computer vision\cite{kleeuse4}, sensor networks\cite{kleever2}, GPUs\cite{kleever1},
device drivers\cite{kleever4}, and online gaming\cite{kleever6}. 
KLEE is also used for automated debugging\cite{kleever3}, thread scheduling\cite{kleever7}, 
and exploit generation\cite{kleever5}. We use KLEE to implement the framework 
proposed in this paper.\par

Siddiqui and Khurshid\cite{juna1,juna2} introduced a novel approach to distributed symbolic execution 
called \emph{ranged analysis}. Their approach involved using a test case to represent a state of the 
execution process. Ranging using tests, referred to as \textit{test ranging}, presents an innovative 
technique to parallelize symbolic execution. This paper builds upon the idea of using test inputs to 
define distinct regions in the program space.  Test ranging allows for using only one type of search 
strategy - depth-first search(DFS) which can be restrictive, especially when trying to maximize coverage 
in a limited amount of time. Having different search strategies also help develop better testing policies. 
This type of ranging also requires full test paths to define boundaries. To address these limitations, we 
introduce an alternative methodology to accomplish ranged analysis called \textit{test-depth partitioning}. 
Our key observation is that the partial paths, also called prefixes, can be used to define non-overlapping 
and exhaustive ranges in the program space and a prefix can be compactly represented as a \emph{test-depth} pair. 
A single test-depth pair represents a pool of execution paths, generated by extending the corresponding prefix. 
These paths are distinct from execution paths made by any other prefix at the same depth.  This inherent 
partitioning can be exploited to define ranges based on these test-depth pairs which can be explored independently 
and using the search technique of choice. A worker executing a region defined by a pair, at any stage, has a set of 
candidate paths that need to be explored. The candidate selection criterion depends on the search strategy. 
When asked to offload work for load balancing, one of the candidate paths is converted to a test and is paired with 
the current depth of the path to make a test-depth couple. This pair represents a prefix which defines a region that 
can be explored by another worker. 
This paper makes the following contributions. 
\begin{itemize}
  \item{\textbf{Test-depth partitioning.}} We use test-depth pairs to represent
               	well-defined regions in program space and develop a method to restrict symbolic
                execution to this region.
  \item{\textbf{Parallel symbolic execution with different search strategies.}} 
								Using \textit{test-depth partitioning},
                we devise a distributed symbolic execution scheme which enables several workers
                to symbolically explore distinct regions of the program space in parallel, 
								employing the search strategy of choice.
  \item{\textbf{Implementation.}} We construct a working prototype of our proposed framework using
                KLEE as the symbolic execution platform.
                \emph{Message passing interface} (MPI)\cite{MPI} is used to implement distributed execution and
                load balancing.
                Out tool is publicly available at:\\
                \hspace*{4ex}{\small\url{https://github.com/Shikhar8990/test-depth-partitioning}}
  \item{\textbf{Evaluation.}} To assess the performance of the proposed framework, we use two sets of 
                applications as our test subjects. The first set comprises twenty-five programs from the GNU Core 
                utilities\cite{coreutil}. 
                The second set consists of three software libraries that perform 
                operations such as processing \emph{YAML} data and implementing an arbitrary precision calculator. 
                An evaluation was carried out to determine the reduction in symbolic execution times of the subjects 
                and analyzing the impact of the factors such as the number of participating workers and the 
                search strategies used.
\end{itemize}

\section{Conceptual Overview} \label{ill}
This section describes the basic symbolic execution process and 
illustrates the core ideas of our framework. The primary objective of symbolic 
execution is to explore all possible 
execution paths in a given program. In contrast to concrete execution 
where variables are assigned explicit values, symbolic execution makes 
one or more input variables symbolic and reasons on the values of these 
variables for different execution paths. 
To illustrate the symbolic exploration procedure and demonstrate 
the functioning of the proposed system, we use a simple \emph{C} 
function that returns the middle integer out of three given integers. 
\begin{figure}[H]
\vspace{-4ex}
\centering
  \begin{lstlisting}
    int find_middle(int x, int y, int z) {
      if (x<y) {
          if (y<z) return y;
          else if (x<z) return z;
          else return x;
      } else if (x<z) return x;
      else if (y<z) return z;
      else return y; 
    } 
\end{lstlisting}
\vspace{-4ex}
\end{figure}

It contains six possible execution paths. Each path is taken under a 
specific value or range of values of the input integers, and these 
restrictions on the inputs generate the path conditions. These path 
conditions can be 
provided to a constraint solver to determine actual values of symbolic 
variables to execute a particular path.  A bit-vector, where a \emph{0} stands 
for a \emph{false} branch while \emph{1} denotes a \emph{true} branch, can represent every 
path in the execution tree. The location of a bit in the bit-vector 
represents the depth of the branch. We refer to such vectors as \textit{test paths}. 
Table~\ref{table:table1} list all test paths in 
the program and their associated constraints.
\begin{table}
  \centering
    \def\arraystretch{1.5}
    \begin{tabular}{ |c|c|c|c| }
      \hline
      \thead{Test} & \thead{Path} & \thead{Constraints} & \thead{Output} \\
      \hline
      \hline
      T1 & 01 & (!(x<y) and (x<z)) & \textit{x} \\
      \hline
      T2 & 11 & ((x<y) and (y<z))  & \textit{y} \\
      \hline
      T3 & 000 & (!(x<y) and !(x<z) and !(y<z)) & \textit{y} \\
      \hline 
      T4 & 001 & (!(x<y) and !(x<z) and (y<z)) & \textit{z} \\
      \hline
      T5 & 100 & ((x<y) and !(y<z) and !(x<z)) & \textit{x} \\
      \hline
      T6 & 101 & ((x<y) and !(y<z) and (x<z)) & \textit{z} \\
      \hline
    \end{tabular}
    \vspace{4pt}
    \caption{Execution paths in \textit{find\_middle}}
    \label{table:table1}
\end{table}

Tests represent full paths as they trace the execution from beginning to end. 
A \textit{prefix} is a partial path that starts at the program entry block but 
does not go all the way to a termination/exit block. The given
program has two feasible prefixes at an execution depth of two \emph{(00, 10)}.
A test-depth pair can be used to represent a prefix; \textit{(T3,2)} and 
\textit{(T4,2)} represent prefix \emph{00} while \textit{(T5,2)} and \textit{(T6,2)} 
represent prefix \emph{10}. Prefixes can become very long for extensive
programs and test-depth pairs can provide a compact representation for partial program paths.
The technique presented in this paper 
uses these test-depth pairs to define non-overlapping and exhaustive regions, 
which can be processed individually as smaller sub-problems.
Symbolic analysis, using a test-depth pair, happens in two phases. 
First, the provided test determines the input value(s) to the program, that are used 
to trace the path taken by the test up to the given depth (provided in the test-depth pair). 
This phase does not 
require any constraint solving as test inputs decide the branching decisions. 
Once execution reaches the assigned depth; 
the second phase begins with a 
symbolic exploration of all possible program paths from that depth level onwards.
In this example, a worker, assigned the pair \textit{(T3/T4,2)}, will end up finding paths 
000 and 001 while another worker, with \textit{(T5/T6,2)}, will find paths 100 and 101. 
Pairs \textit{(T1,2)} and \textit{(T2,2)} represent test paths and can not be extended further.\par
\begin{figure}
  \centering
  \subfloat[Execution tree of \textit{find\_middle}]
    { \includegraphics[width=.4\linewidth]{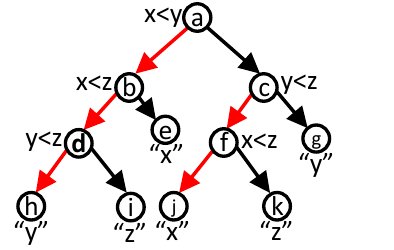}
      \label{fig:tree}
    }
  \subfloat[Worker execution tree]{
    \includegraphics[width=.4\linewidth]{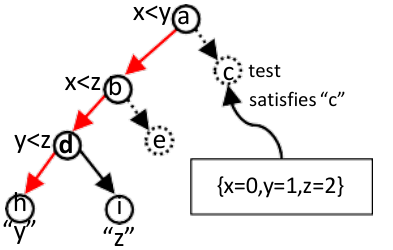}
    \label{fig:wtree}
  }
  \caption{Test-depth ranging}
  \label{fig:illustration}
\end{figure}
Figure~\ref{fig:illustration}a depicts the execution tree of the \textit{find_middle} subroutine.
Red and black arrows indicate false and true branches respectively.
As has already been discussed, the code contains six full paths which produces six tests.  
The symbolic execution process starts from the first line of the program and generates an 
initial state \textbf{a}. When execution hits the first branch, two states - \textbf{b} 
and \textbf{c}, representing the \emph{if} and the \emph{fall-through} branches are generated. 
Subsequent states labeled \textbf{d} to \textbf{k} are spawned through the course of execution. 
Each state keeps a track of the path constraints required to trace the path it represents. 
Let us assume a scenario where a worker starts execution with a test-depth pair 
of \emph{\{\{x=1,y=0,z=0\},2\}}. Figure~\ref{fig:illustration}b shows the execution path of this worker. It relies
on the test inputs to guide its execution for the first two branches as the given depth 
in the pair is \emph{two}. On the first branch, it will only explore the \textit{false} branch and suspend 
state \textbf{c} which is not satisfied by the test. Similarly, the second branch would 
only add state \textbf{d} while suspending \textbf{e}. The execution sub-trees beginning at the suspended 
states are not explored. When this worker receives another test-depth pair - \emph{\{\{x=0,y=1,z=2\},2\}}, 
the suspended states are searched to find a state whose constraints are satisfied by the new test. 
In this case, the test inputs satisfy state \textbf{c}, and it gets activated for further exploration. 
Since the provided depth is two, the test would have guided the next branch as well, 
adding state \textbf{g} and suspending \textbf{f}.

\section{Technique}
\subsection{Bounding program space using test-depth pairs}
\begin{algorithm}
	\caption{Restricting execution range using a test-depth pair}
  \textbf{input:} \text{current\_state, ($\tau$,$\tau\_depth$)}\\
	\textbf{output:} \text{states to explore at next depth}
  \begin{algorithmic}[1] 
		\State $cond \leftarrow \text{branch condition}$
		\State $curr\_depth \leftarrow \text{depth of current state}$
		\State $BB\_true \leftarrow \text{state for true branch}$
		\State $BB\_false \leftarrow \text{state for false branch}$
    \If{$curr\_depth < \tau\_depth$}
      \Comment{Restrict the path to follow the test}
      \State $branchTaken \leftarrow \Call{solvePath}{\tau, cond}$
			\If{$branchTaken \text{ is true}$}
        \State $\Call{suspendState}{BB\_false}$
				\State $\textbf{return } BB\_true$
			\Else
        \State $\Call{suspendState}{BB\_true}$
				\State $\textbf{return } BB\_false$
			\EndIf
		\Else
      \Comment{Execution reached the assigned region}
			\If{$cond \text{ is satisfiable}$}
				\State $\textbf{return } BB\_true$
			\ElsIf{$\neg cond \text{ is satisfiable}$}
				\State $\textbf{return } BB\_false$
			\Else
				\State $\textbf{return } (BB\_false,BB\_true)$	
			\EndIf
		\EndIf		
	\end{algorithmic}
  \label{algo1}
\end{algorithm}

To accomplish test-depth partitioning, we modify the standard symbolic execution 
algorithm to allow for a test to guide the execution up to a certain depth and 
then explore all possible paths below that depth. Algorithm~\ref{algo1} describes 
this augmentation. The exploration process represents the program paths as execution 
states. The execution on reaching a branch generates states to capture 
both the \emph{true} and \emph{false} branches ($BB\_true \text{ and } BB\_false$)(lines 3-4). 
When restricting the 
program space using a test-depth pair ($\tau, \tau\_depth$), on hitting a branch, 
one of two things can happen. If the execution depth is less than test depth 
($\tau\_depth$), we let the test inputs ($\tau$) decide the branch and suspend the 
state corresponding to the \emph{not-taken} branch (lines 6-12). The function $SOLVEPATH$ assigns the test 
inputs to the branch condition and determines the branching decision.
Since a concrete test will always take 
only one branch, this ensures only one specific path gets replayed till $\tau\_depth$. 
Standard symbolic execution resumes once execution reaches the assigned region 
(deeper than $\tau\_depth$) and satisfiability of branch conditions ($cond$) determines 
path exploration (lines 14-19).
\subsection{Distributed symbolic execution}
This section discusses the distributed symbolic execution framework to enable different 
workers to explore non-overlapping parts of the program space. The setup comprises a 
coordinator node and several worker nodes. The coordinator is responsible for generating 
a pool of test-depth pairs and assigning them to the worker nodes. Each worker symbolically 
executes the paths in region of the assigned pair. The workers have 
the freedom to decide the search strategy. This method also overcomes a major 
limitation of previous work on ranging, which could only use depth-first search. Load 
balancing happens through work stealing. 
Workers can quickly convert an unexplored path into a test-depth pair and send it
to the coordinator which sends it to the idle worker.
\subsubsection{Coordinator operation}
\begin{algorithm}
  \caption{Coordinator operation}
  \begin{algorithmic}[1]
    \LineComment{Initialization}
    \State $numWorkers \leftarrow \text{number of worker nodes in addition to the coordinator}$
    \newline
    \LineComment{Initially all workers are idle}
		\State $free\_List \leftarrow \text{list of worker nodes}$
		\State $busy\_List \leftarrow \text{empty list}$ 
    \newline
    \LineComment{workers will terminate when all paths reach this depth}
    \State $final\_depth \leftarrow \text{upper bound on exploration depth}$
    \newline
    \LineComment{Search options like depth-first, breadth-first or random search}
    \State $searchType \leftarrow \text{search strategy to use}$
    \newline
    \newline
    \Comment{Generate initial tests for each worker}
    \State $test\_depth\_pairs \leftarrow \Call{executeSymbolic}{searchType, numWorkers}$
    \newline
    \newline
    \Comment{Seeding each worker with a test and starting distributed execution}
    \For {\text{$worker$ in $free\_List $}}
      \LineComment{Distribute the initial tests and update free and busy lists}
      \State $(\tau, \tau\_depth) \leftarrow \Call{pickAtest}{test\_depth\_pairs}$
      \State $\Call{SEND}{worker,searchType,\tau,\tau\_depth,final\_depth}$
      \State $\Call{removeFrom}{free\_List,worker}$
      \State $\Call{addTo}{busy\_List,worker}$
    \EndFor
    \newline

    \LineComment{Start Load Balancing using work stealing}
    \While{true}
      \State $(task, worker) \leftarrow \Call{recv}$
      \Comment{Receive message from workers}
      \If{task = \text{finish}}
        \State $\Call{removeFrom}{busy\_List,worker}$
        \State $\Call{addTo}{free\_List,worker}$
        \If{$free\_List.size() = numWorkers$}
          \LineComment{Execution complete, terminate all workers}
          \State $\Call{SEND}{all\_workers, terminate}$
          \State $return$
        \EndIf
      \EndIf
      \newline
      \If{$free\_List.size() > 0$} \Comment{One or more workers are idle}
        \State $idle\_Worker \leftarrow \Call{pickAworker}{free\_List}$ 
        \State $busy\_Worker \leftarrow \Call{pickAworker}{busy\_List}$
        \State $\Call{Send}{busy\_Worker, provide\_work}$
        \State $((\tau, \tau\_depth), busy\_Worker) \leftarrow \Call{recv}$
        \If{$\tau$ is valid}
          \State $\Call{Send}{idle\_Worker, searchType, \tau, \tau\_depth, final\_depth}$
          \State $\Call{removeFrom}{free\_List,idle\_Worker}$
          \State $\Call{addTo}{busy\_List,idle\_Worker}$
        \Else
          \State \text{work stealing failed; try again}
        \EndIf
      \EndIf 
    \EndWhile
	\end{algorithmic}
  \label{algo2}
\end{algorithm}

The coordinator node is responsible for assigning and managing work amongst worker nodes. 
Algorithm~\ref{algo2} shows a simplified version of the tasks performed by the coordinator. 
Initially, it undertakes a shallow symbolic execution to generate an initial pool of 
test-depth pairs (line 6) to distribute. The current configuration assigns one pair to each 
worker (lines 7-11). The coordinator also keeps track of busy and idles workers 
($free\_List, busy\_List$). After the initial distribution of work, it monitors messages 
from the worker nodes (line 13). If a worker finishes exploring its assigned region (line 14), 
the \emph{busy} and \emph{idle} lists get updated (lines 15-16).  The search terminates when all workers 
become idle (line 18). In case a worker becomes available, the coordinator asks one of the 
busy workers to provide work to the idle worker in the form of another test-depth pair 
(lines 21-24).  The configuration shown explores up to a finite depth ($final\_depth$). 
However, the framework also supports time bounded exploration. The search policy determines
the \emph{searchType}.
\subsubsection{Worker operation}
\begin{algorithm}
	\caption{Worker operation}
  \begin{algorithmic}[1]
    \LineComment{Each worker maintains a list of active and suspended states}
    \State $suspendedStates \leftarrow \text{list of suspended states}$
    \State $activeStates \leftarrow \text{list of active states}$
    \newline
    \LineComment{Main execution loop}
    \While{true}
      \State $task \leftarrow \Call{recv}$
      \Comment{Receive message from co-ordinator}
      \If{$task=(searchType,\tau,\tau\_depth,final\_depth)$}
        \If{$\text{suspendedStates is empty}$}
          \Comment{First test, start execution from the beginning}
          \State $initialState \leftarrow \Call{generateInitialState}$
          \State $\Call{addToActiveStates}{initialState}$
          \State $\Call{StartExecution}{initialState, \tau, \tau\_depth, final\_depth}$
          \Comment{Start executing\newline \hspace*{23em}from this state following $\tau$ up to $\tau\_depth$}
          \newline
          \State $\Call{Send}{\text{finish}}$ \Comment{notify the coordinator after finishing}
        \Else
          \LineComment{Look for a suspended state satisfying the test and resume it}
          \For{\text{$state$ in $suspendedStates$}}
            \State $contraints \leftarrow \Call{getConstraints}{state}$
            \State $sat \leftarrow \Call{solveConstraints}{\tau,contraints}$
            \If{$sat \text{ is }true$}
              \Comment{Found a state to resume}
              \State $\Call{addTo}{activeStates,state}$
              \State $\Call{removeFrom}{suspendedStates,state}$
              \State $\Call{StartExecution}{state, \tau, \tau\_depth, final\_depth}$
              \State $\Call{Send}{\text{finish}}$
              \State $break;$
            \EndIf
          \EndFor
        \EndIf
      \EndIf
      \newline
      \LineComment{Send work to the coordinator which will send it to an idle worker}
      \If{$task=provide work$}
        \If{$activeStates.size() > offloadThreshold$}
          \State $state \leftarrow \Call{pickastate}{activeStates}$
          \State $\tau \leftarrow \Call{createTest}{state}$
          \State $\tau\_depth \leftarrow \Call{getDepth}{state}$
          \LineComment{Send the test-depth pair to the coordinator}
          \State $\Call{Send}{\tau, \tau\_depth}$
        \Else
          \LineComment{No work to give}
          \State $\Call{Send}{\text{no_work}}$
        \EndIf
      \EndIf
      \If{$task=terminate$}
        \LineComment{Exploration Complete}
        \State $break;$
      \EndIf
    \EndWhile
	\end{algorithmic}
  \label{algo3}
\end{algorithm}

Algorithm~\ref{algo3} describes the various operations carried out by the workers.
These nodes, responsible for carrying out symbolic execution in their assigned regions, 
maintain a list of states that are active and can be explored further and states that are 
suspended. When a worker receives a new task from the coordinator (line 5), it can start from 
the program entry point if this is the first test-depth pair (lines 7-9), or it can resume 
one of the suspended states, whose constraints are satisfied by $\tau$ (lines 13-14), and 
explore it further (lines 16-18). The worker node notifies the coordinator after completion 
of the assigned task (lines 10, 19). 
Workers can be asked to offload some work for load balancing. If it has enough active 
states (line 22), it can choose one of the states (line 23), solve its path constraints to create 
a test, combine it with the depth of the state to create a test-depth pair (lines 24-25) and send 
it to the coordinator (line 26). Work stealing fails in case there are not sufficient active 
states (line 28). In the present implementation, a worker will only offload a test-depth pair if it
has more the four active states (\textit{offloadThreshold} is 4). However, this value can be easily 
modified according to the requirements.
\subsection{Implementation}
KLEE\cite{klee1} is a symbolic execution tool which works on the LLVM\cite{llvm} compiler 
(IR) intermediate representation. Symbolic execution on compiler IR is an attractive proposition. 
The representation is closer to the machine, and the branching structure is simplified, 
allowing for only two branch targets. We augment the KLEE source code to implement 
\emph{test-depth partitioning}. 
Distributed execution and load balancing are implemented using MPI\cite{MPI}.

\section{Evaluation}
\subsection{Benchmarks}
The proposed framework is evaluated using GNU Coreutils (version 6.11)\cite{coreutil} 
and three libraries. Coreutils are one of the most heavily used tools in 
Unix-like systems and deal with shell, text, and file system operations. 
We use twenty-five programs with the largest LLVM bit-code files from this 
suite.  The second set of subjects comprise the GNU oSIP 4.0.0\cite{osip}, 
GNU BC 2.27\cite{bc}, and LibYAML 0.1.5\cite{yaml} libraries. KLEE has to process 
multiple source files to analyze these applications.
oSIP implements the Session 
Initiation Protocol (SIP) and provides an API for applications to incorporate 
the protocol. BC is used to perform arbitrary precision numeric processing. 
Its syntax is similar to C, and it can be used either as an interactive 
calculator or as a scripting language. LibYAML provides the functionality to 
parse and process data in YAML format, which is a serialization format designed 
to be human-readable. The execution drivers for these libraries are adapted from \cite{chopper}.
\subsection{Methodology}
To demonstrate the efficacy of our scheme, we observe the speed-ups achieved 
when using multiple workers. For a defined program space, we compare the time 
taken to complete the symbolic execution when using one worker with the execution 
times when using more than one worker. We evaluate two distributed 
configurations - one comprising two worker nodes and the other comprising 
four workers. For a given test application, the exploration space has to be consistent 
across configurations.  To accomplish this, we run a breadth-first search(BFS) for 
all the programs with a timeout of six-hundred seconds using one worker.
The depth to which symbolic execution reaches before timeout provides the exploration
upper-bound.
For subsequent experiments analyzing the run-times, execution terminates when all states
across all workers reach this depth which acts as a termination criterion.
This results in a well-defined bounded exploration space.
To make the evaluation comprehensive, we use three exploration strategies - 
depth-first, breadth-first, and random-state search.  The search technique determines 
how states are explored. Depth-first strategy, upon reaching a branching point, forks 
execution and follows one branch till a termination point and then backtracks, 
generating deeper states in the process. Breadth-first search processes all states 
at a particular depth before proceeding to the next depth. Random-state search, as 
the name suggests, randomly picks a state to execute from the pool of active states.
KLEE provides DFS and random-state search capabilities but lacks support for BFS. 
We developed a custom breadth-first search option to enable experimentation. 
Table~\ref{table:table2} lists details of the hardware platform and software versions used 
for carrying out the experiments. 
\begin{table}
  \centering
  \def\arraystretch{1.5}
  \begin{tabular}{ |c|c| }
    \hline
    CPU & Intel(R) Core(TM) i7-8700K\\
    \hline
    \#Cores & 12\\
    \hline
    Memory & 32Gb\\
    \hline
    OS & Ubuntu 16.04 LTS\\
    \hline
    Klee Version & 3.4 (commit SHA - d2fbdf7)\\
    \hline
    LLVM Version & 3.4\\
    \hline
    SMT solver & STP 2.3.3 (commit SHA - 03aa904)\\
    \hline
    MPI Version & Open MPI 1.10.2\\
    \hline
  \end{tabular}
  \vspace{4pt}                                              
  \caption{System configuration}
  \label{table:table2}         
\end{table}

\subsection{Results}
This section discusses the results of experiments aimed to determine the speed-ups 
achieved when exploring a defined region of program space using test-depth partitioning. 
As already mentioned, we use twenty-five programs from GNU Coreutils and three software 
libraries. Table~\ref{table:table3} list these applications, the depth of exploration, 
and the total paths explored. We also list the number of test transfers that take place
due to load balancing, when using a 
four-worker configuration.  The resulting speed-ups discussed in the following sections 
are a result of the parallelism gained by a simultaneous exploration of different parts 
of the execution tree. The primary overhead of a distributed scheme like ours is that 
of \textit{test-depth pair} transfers between workers during load balancing. The benefit of such 
a transfer depends on the size of the execution tree it represents.  A test-depth transfer 
is beneficial if the communication and processing overhead is more than compensated by 
the parallelism gained. Transferring a test-depth pair, which represents an ample 
execution space to explore is more advantageous than a pair which terminates at a shallow 
depth. A significant challenge is that it is tough to make an apriori assessment of the 
size of the execution space represented by a test-depth pair. To alleviate this issue, in 
our scheme, when a worker is asked to offload one unexplored path to another worker, 
it gives preference to active states at shallow depths. A state near to the root of the 
tree is likely to represent a larger space than one farther away from it.
Another factor impacting the observed speed-ups is the constraint-caching mechanism that 
KLEE uses where queries sent to the SMT solver are cached and reused for other paths. 
Calls to the solver are primary contributors to the execution times, and caching eliminates 
redundant SMT solving. Partitioning scheme like ours can also impact cache hit rates. 
Caching might not be very useful for a worker executing a series of test-depth pairs that 
represent paths lying in very different regions of the program space while it will benefit a 
worker executing on regions in close proximity. Caching can also be impacted by the 
search technique used.
The overall performance of test-depth partitioning depends on the interplay between these factors 
which vary from application to application.
\begin{table}
  \centering
  \def\arraystretch{1.5}
    \begin{tabular}{ |p{2cm}|p{1.5cm}|p{1.5cm}|p{1.5cm}|p{1.5cm}|p{1.5cm}| }
      \hline
      {} & {} & {} & \multicolumn{3}{p{4.5cm}|}{\thead{Test transfers}}\\
      \hline
      \thead{Application} & \thead{Depth} & \thead{Paths} & \thead{DFS} & \thead{Random} &\thead{BFS}\\
      \hline
      \hline
      dir & 13 & 3585 & 12 & 487 & 874\\
      \hline
      ls & 13 & 3585 & 19 & 125 & 870\\
      \hline
      dd & 16 & 7265 & 19 & 1250 & 732\\
      \hline
      join & 16 & 8545 & 13 & 1903 & 3338\\
      \hline
      mkdir & 16 & 9105 & 14 & 1808 & 853\\
      \hline
      chcon & 17 & 14859 & 17 & 3125 & 0\\
      \hline
      chgrp & 17 & 14795 & 16 & 4331 & 460\\
      \hline
      chown & 17 & 14795 & 13 & 3138 & 466\\
      \hline
      cp & 17 & 14757 & 12 & 3144 & 4015\\
      \hline
      date & 17 & 14621 & 19 & 2902 & 5454\\
      \hline
      ginstall & 17 & 14700 & 21 & 5410 & 4298\\
      \hline
      mv & 17 & 14859 & 17 & 4353 & 458\\
      \hline
      rm & 17 & 18375 & 17 & 7571 & 7883\\
      \hline
      nl & 20 & 22578 & 46 & 10141 & 5306\\
      \hline
      ptx & 21 & 24941 & 19 & 8993 & 10217\\
      \hline
      pr & 22 & 77047 & 19 & 23527 & 5303\\
      \hline
      ln & 23 & 160644 & 28 & 51512 & 25007\\
      \hline
      tac & 24 & 32996 & 29 & 14581 & 9173\\
      \hline
      expr & 27 & 54713 & 22 & 16259 & 8157\\
      \hline
      sha512sum & 31 & 284727 & 72 & 60253 & 31827\\
      \hline
      head & 25 & 126151 & 32 & 28256 & 28912\\
      \hline
      printf & 18 & 27391 & 17 & 5170 & 9894\\
      \hline
      readlink & 27 & 168229 & 38 & 48996 & 29555\\
      \hline
      split & 23 & 50506 & 36 & 27710 & 19233\\
      \hline
      sum & 29 & 23707 & 16 & 10729 & 0\\
      \hline
      \hline
      bc & 27 & 89927 & 36 & 23138 & 22100\\
      \hline
      osip & 38 & 71917 & 28 & 25685 & 5495\\
      \hline
      libyaml & 23 & 22811 & 45 & 8902 & 2644\\
      \hline
    \end{tabular}
    \vspace{4pt}
    \caption{Paths explored and tests exchanged with 4 workers}
    \label{table:table3}
\end{table}

\begin{figure*}
\centering
%\begin{figure*}
  %\scriptsize
  %\centering
  \subfloat[depth-first search] {
  \begin{tikzpicture}
  \begin{axis} [only marks, xmin=dir, xmax=sum,
      width=\textwidth, xlabel near ticks, ylabel near ticks,
      height=5cm, xmajorgrids=true, ymajorgrids=true,
      xlabel=Application, ylabel=Speed-up, ymin=0, ymax=6, ytick={0,1,2,3,4,5,6},
            symbolic x coords={dir,ls,dd,join,mkdir,chcon,chgrp,
            chown,cp,date,ginstall,mv,rm,nl,ptx,pr,ln,tac,expr,
            sha512sum,head,printf,readlink,split,sum},xtick=data, xticklabel style={rotate=90}] 
  \addplot+[mark=*, color=green,mark options=solid] plot coordinates{
    (dir,1.820197014)
    (ls,0.9179162411)
    (dd,2.425356688)
    (join,2.406915521)
    (mkdir,2.485376054)
    (chcon,2.550746936)
    (chgrp,2.243704136)
    (chown,2.213598242)
    (cp,2.805182501)
    (date,2.727337834)
    (ginstall,2.301285017)
    (mv,2.268361373)
    (rm,1.952853998)
    (nl,2.086930868)
    (ptx,1.662367234)
    (pr,1.891538734)
    (ln,1.652562766)
    (tac,1.79346264)
    (expr,1.613595162)
    (sha512sum,1.729017816)
    (head,1.693764696)
    (printf,2.150115198)
    (readlink,1.763648521)
    (split,1.841775229)
    (sum,1.580819892)
  };
  \addlegendentry{2 workers}
  \addplot+[mark=*, color=blue,mark options=solid] plot coordinates{
    (dir,3.296301462)
    (ls,1.789966707)
    (dd,4.211001558)
    (join,4.622281201)
    (mkdir,4.683917774)
    (chcon,4.811585099)
    (chgrp,4.356975144)
    (chown,4.425527556)
    (cp,5.692140244)
    (date,4.811196726)
    (ginstall,4.520148213)
    (mv,4.410449546)
    (rm,3.666227587)
    (nl,3.734551255)
    (ptx,2.977181533)
    (pr,3.297065781)
    (ln,3.044569137)
    (tac,2.94395084)
    (expr,2.82530117)
    (sha512sum,3.116661323)
    (head,2.764715991)
    (printf,3.91690842)
    (readlink,3.242102258)
    (split,2.779067045)
    (sum,2.593642529)
  };
  \addlegendentry{4 workers}
  \end{axis}
  \end{tikzpicture}}
  %\caption{Speed-up using depth-first search}
  %\label{speedupdfs}
%\end{figure*}
\newline
%\begin{figure*}
  %\scriptsize
  %\centering
  \subfloat[random-state search] {
  \begin{tikzpicture}
  \begin{axis} [only marks, xmin=dir, xmax=sum,
      width=\textwidth, xlabel near ticks, ylabel near ticks,
      height=5cm, xmajorgrids=true, ymajorgrids=true,
      xlabel=Application, ylabel=Speed-up, ymin=0, ymax=6, ytick={0,1,2,3,4,5,6},
            symbolic x coords={dir,ls,dd,join,mkdir,chcon,chgrp,
            chown,cp,date,ginstall,mv,rm,nl,ptx,pr,ln,tac,expr,
            sha512sum,head,printf,readlink,split,sum},xtick=data, xticklabel style={rotate=90}] 
  \addplot+[mark=*, color=green,mark options=solid] plot coordinates{
    (dir,1.927688313)
    (ls,2.086061548)
    (dd,2.045142211)
    (join,2.043881128)
    (mkdir,2.302837962)
    (chcon,2.277362931)
    (chgrp,2.21567116)
    (chown,2.244014817)
    (cp,2.277880871)
    (date,2.273336998)
    (ginstall,2.275685127)
    (mv,2.338250749)
    (rm,2.292709272)
    (nl,1.596644966)
    (ptx,3.893842096)
    (pr,2.242563995)
    (ln,1.929898084)
    (tac,1.617021034)
    (expr,2.119541259)
    (sha512sum,1.922684386)
    (head,1.303527402)
    (printf,1.648241304)
    (readlink,1.162828917)
    (split,1.598316453)
    (sum,1.454673182)
  };
  \addlegendentry{2 workers}
  \addplot+[mark=*, color=blue,mark options=solid] plot coordinates{
    (dir,4.323047306)
    (ls,3.134749194)
    (dd,3.492507711)
    (join,3.452952283)
    (mkdir,4.038689477)
    (chcon,3.772434462)
    (chgrp,3.372074274)
    (chown,3.734886575)
    (cp,3.995199462)
    (date,4.051888404)
    (ginstall,3.562548347)
    (mv,3.885674712)
    (rm,5.145475136)
    (nl,3.17416187)
    %(ptx,8.888427394)
    (ptx,6.0)
    (pr,4.781974057)
    (ln,2.791218054)
    (tac,2.702394128)
    (expr,3.834536645)
    (sha512sum,2.020292405)
    (head,2.155594697)
    (printf,3.403563755)
    (readlink,1.387628368)
    (split,2.373150985)
    (sum,1.846608661)
  };
  \node at (axis cs:ptx,5.5) {>6x};
  \addlegendentry{4 workers}
  \end{axis}
\end{tikzpicture}}
  %\caption{Speed-up using random search}
  %\label{speeduprand}              
%\end{figure*}                     
\newline
%\begin{figure*}
  %\scriptsize
  %\centering
\subfloat[breadth-first search] {
  \begin{tikzpicture}
  \begin{axis} [only marks, xmin=dir, xmax=sum,
      width=\textwidth, xlabel near ticks, ylabel near ticks,
      height=5cm, xmajorgrids=true, ymajorgrids=true,
      xlabel=Application, ylabel=Speed-up, ymin=0, ymax=6, ytick={0,1,2,3,4,5,6},
            symbolic x coords={dir,ls,dd,join,mkdir,chcon,chgrp,
            chown,cp,date,ginstall,mv,rm,nl,ptx,pr,ln,tac,expr,
            sha512sum,head,printf,readlink,split,sum},xtick=data, xticklabel style={rotate=90}] 
  \addplot+[mark=*, color=green,mark options=solid] plot coordinates{
    (dir,1.839869974) 
    (ls,1.907289709) 
    (dd,1.80797543) 
    (join,1.816333583) 
    (mkdir,1.889415974) 
    (chcon,2.035485349) 
    (chgrp,2.037894682) 
    (chown,2.046774779) 
    (cp,2.126496278) 
    (date,1.915493186) 
    (ginstall,2.031419479) 
    (mv,2.03454965) 
    (rm,1.816208244) 
    (nl,1.341293962) 
    (ptx,1.525643555) 
    (pr,1.160960494) 
    (ln,1.313583468) 
    (tac,1.4034101) 
    (expr,1.195767388) 
    (sha512sum,1.6) 
    (head,1.4) 
    (printf,1.511115309) 
    (readlink,1.067899216) 
    (split,1.062534574) 
    (sum,1.174412098)
  };
  \addlegendentry{2 workers}
  \addplot+[mark=*, color=blue,mark options=solid] plot coordinates{
    (dir,1.119957123)
    (ls,1.09055327)
    (dd,2.673402221)
    (join,2.922413646)
    (mkdir,2.834139187)
    (chcon,3.068359704)
    (chgrp,3.00744287)
    (chown,3.001345215)
    (cp,3.41738627)
    (date,3.02462194)
    (ginstall,3.237183327)
    (mv,2.999989165)
    (rm,3.441276786)
    (nl,2.020172337)
    (ptx,2.85292867)
    (pr,1.22127013)
    (ln,1.979749226)
    (tac,2.011714286)
    (expr,1.74528288)
    (sha512sum,2.9)
    (head,2.8)
    (printf,2.370181818)
    (readlink,1.360466432)
    (split,1.24332446)
    (sum,1.938326389)
  };
  \addlegendentry{4 workers}
  \end{axis}
  \end{tikzpicture}
}
  %\caption{Speed-up using breadth-first search}
  %\label{speedupbfs}
%\end{figure*}
  \caption{Test-depth partitioning performance for Coreutils applications}
\label{speed}
\end{figure*}
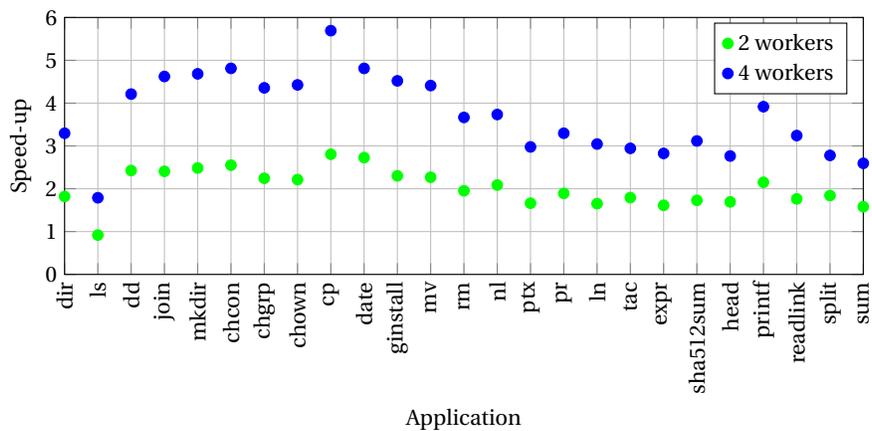
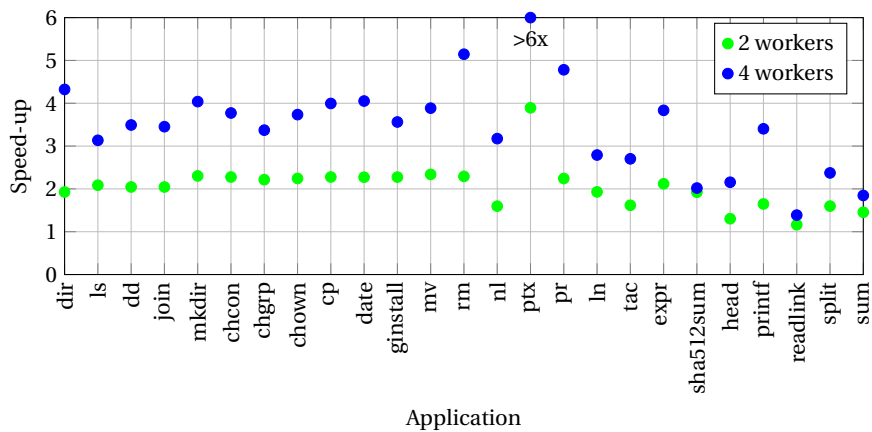
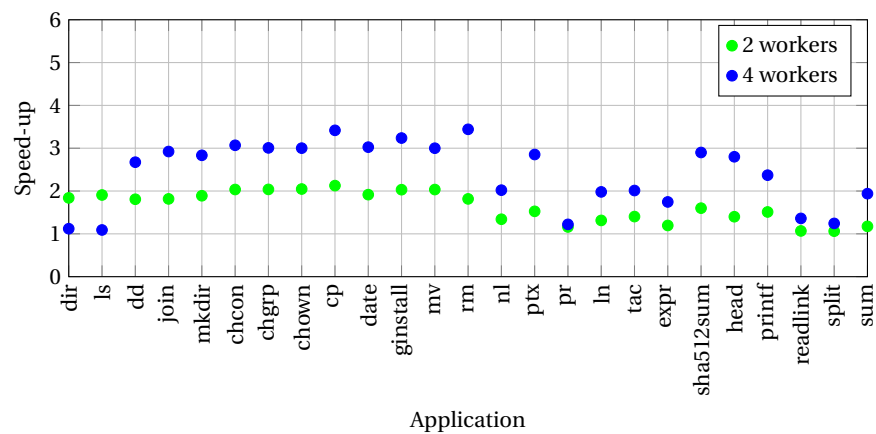

\subsubsection{Coreutils}
Figure~\ref{speed} shows the speed-ups delivered across two different configurations 
comprising two and four workers and three search techniques. 
Depth-first and random-state searches achieve median linear speed-ups when using 
two workers and an almost linear speed-up of \emph{3.7x} and \emph{3.6x} respectively 
when using four workers. 
These results indicate narrow execution trees for these programs. 
Using four workers and DFS for program \emph{cp} 
results in a fairly balanced partitioning with only a few test-depth transfers (16) due to 
load balancing, which results in a speed-up of almost \emph{6x}. This application also witnesses 
a linear speed-up with random-state search. However, using BFS decreases the speed-up 
to \emph{3.4x}. The program \emph{ptx} achieves a speed-up of more than \emph{6x} with four 
workers and a random-state search. 
Eleven programs with DFS and seven programs with a random-state search see more than 
\emph{4x} speed-up when executed with four workers. With two workers, twelve and fourteen programs 
see super-linear improvements with DFS and random search respectively.  
A few applications like \emph{ls}, \emph{sum} and \emph{split} do not benefit a lot from parallel 
execution. In these cases, the benefits of distributing work are overshadowed by overheads 
associated with inter-worker communication and a possible reduction in constraint caching performance.  
In fact, for \emph{dir} and \emph{ls}, the execution times increase when going from two to four workers 
due to these overheads. Coreutils programs also display a marked diversity in behavior with speed-ups 
ranging from \emph{1x} to more than \emph{6x} across different configurations. 
\subsubsection{Libraries}
\begin{figure*}
  %\scriptsize
  \centering
	\subfloat[depth-first search] {
		\begin{tikzpicture}
			\begin{axis} [only marks, xmin=bc, xmax=libyaml,
					width=4cm, xlabel near ticks, ylabel near ticks,
					height=5cm, xmajorgrids=true, ymajorgrids=true,
          ymin=0, ymax=8, ytick={1,2,3,4,5,6,7,8},
					xlabel=Application, ylabel=Speed-up,
          legend style={at={(0.15,0.85)},anchor=west},
								symbolic x coords={bc, osip, libyaml},xtick=data, xticklabel style={rotate=90}] 
			\addplot+[mark=*, color=green,mark options=solid] plot coordinates{
				(bc,1.682307019)
				(osip,1.677693311)
				(libyaml,3.08683353)
			};
			\addlegendentry{2 workers}
			\addplot+[mark=*, color=blue,mark options=solid] plot coordinates{
				(bc,3.143302544)
				(osip,3.13627146)
				(libyaml,7.122849553)
			};
			\addlegendentry{4 workers}
			\end{axis}
		\end{tikzpicture}
	}
	\subfloat[random state] {
		%\scriptsize
		\centering
		\begin{tikzpicture}
		\begin{axis} [only marks, xmin=bc, xmax=libyaml,
				width=4cm, xlabel near ticks, ylabel near ticks,
				height=5cm, xmajorgrids=true, ymajorgrids=true,
        ymin=0, ymax=8, ytick={1,2,3,4,5,6,7,8},
				xlabel=Application, ylabel=Speed-up,
        legend style={at={(0.15,0.85)},anchor=west},
							symbolic x coords={bc, osip, libyaml},xtick=data, xticklabel style={rotate=90}]
		\addplot+[mark=*, color=green,mark options=solid] plot coordinates{
			(bc,2.82)
			(osip,1.27)
			(libyaml,1.73)
		};
		\addlegendentry{2 workers}
		\addplot+[mark=*, color=blue,mark options=solid] plot coordinates{
			(bc,7.02)
			(osip,1.75)
			(libyaml,3.85)
		};
		\addlegendentry{4 workers}
		\end{axis}
		\end{tikzpicture}
	}
	\subfloat[breadth-first search] {
		\centering
		\begin{tikzpicture}
		\begin{axis} [only marks, xmin=bc, xmax=libyaml,
				width=4cm, xlabel near ticks, ylabel near ticks,
				height=5cm, xmajorgrids=true, ymajorgrids=true,
        ymin=0, ymax=8, ytick={1,2,3,4,5,6,7,8},
				xlabel=Application, ylabel=Speed-up,
        legend style={at={(0.15,0.85)},anchor=west},
							symbolic x coords={bc, osip, libyaml},xtick=data, xticklabel style={rotate=90}]
		\addplot+[mark=*, color=green,mark options=solid] plot coordinates{
			(bc,1.2)
			(osip,1.33)
			(libyaml,1.72)
		};
		\addlegendentry{2 workers}
		\addplot+[mark=*, color=blue,mark options=solid] plot coordinates{
			(bc,2)
			(osip,3.3)
			(libyaml,3.2)
		};
		\addlegendentry{4 workers}
		\end{axis}
		\end{tikzpicture}
	}
\caption{Test-depth partitioning performance for \emph{bc, osip} and \emph{libYAML}}
\label{speedlib}
\end{figure*}
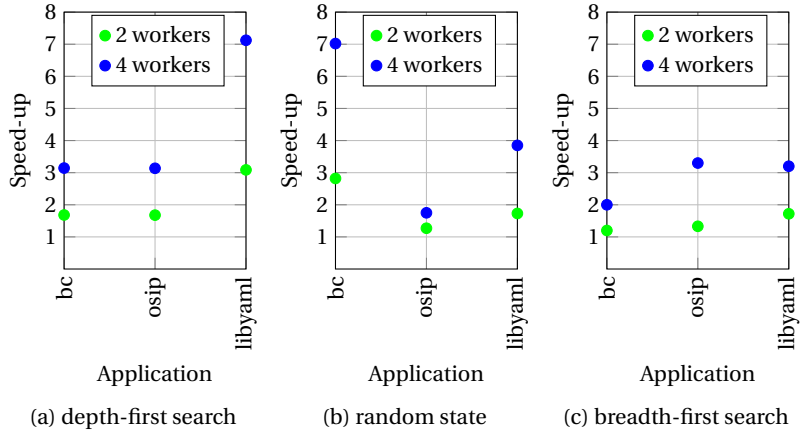

The three application libraries are analyzed using a methodology similar to the one used for 
Coreutils programs. Figure~\ref{speedlib} depicts the speed-ups achieved across three search 
techniques and two configurations. Depth-first search provides a median speed-up of \emph{1.7x} 
and \emph{3.2x} for two and four worker configurations, respectively. 
Random-state performs similar to DFS for the two core configurations while improving the four 
worker median speed-up to \emph{3.85x} for the three subjects. Breadth-first search exhibits median 
speed-ups of \emph{1.33x} with two workers and \emph{3.2x} for four workers. A key observation is that 
the one search strategy might not always be the best across the board. For eg., \emph{libYAML}, 
with four workers employing DFS, sees a \emph{7x} speed-up while \emph{bc}, with the same configuration, 
sees only a \emph{3.13x} speed-up. However, if the search strategy changes to random-state, 
\emph{bc} witnesses a higher speed-up (\emph{7x}) than \emph{libYAML}(\emph{3.85x}). oSIP benefits the least
from our partitioning and distributing scheme with a maximum speed-up of \emph{3.3x} with four workers using 
BFS.

\section{Related Work}
There have been several approaches to enable distributed symbolic execution.
The current work builds upon the concept of ranged symbolic execution 
\cite{juna1,juna2}, which partitions the program space using tests. 
In the original work, a total ordering on tests is used to define ranges 
which start at a particular test $\tau$1 and end at a test $\tau$2 where 
$\tau$2>$\tau$1 according to the ordering. A range comprises paths that 
lie between the two bounding tests. This idea is extended to devise a 
distributed approach where different workers explore ranges defined by a 
pair of tests. Load balancing is accomplished using work-stealing. 
When a worker, performing execution on a given range, hits a branch, 
it explores the \emph{true} branch puts the \emph{false} one in a queue. Other workers can 
steal from this queue.  If a node receives a request to offload work, 
one of the states in the queue is converted to a test case and passed to the 
coordinator, and its current range is modified to end at that test. For example, 
when a worker, executing a region between tests $\tau1$ and $\tau2$, receives 
a request to offload work. It uses one of its \textit{false} branch states in 
the queue to a construct a test case $\tau$3; it offloads the range between $\tau$3 
and $\tau2$ while its range updates to one between $\tau1$ and $\tau3$. Non-overlapping 
and exhaustive ranges are guaranteed as long as all the workers explore the \emph{true} 
branch before the \emph{false} branch, which enforces a depth-first strategy. Test-depth 
partitioning overcomes this limitation by incorporating the \text{depth} of a state 
to enable distribution of partial paths.
Cloud9\cite{cloud9}, a well known framework, defines region boundaries 
in terms of \textit{fence nodes} and the pool of states to be explored comprise the \textit{candidate nodes}. 
Participating workers get a part of the execution tree, and work transfers involve sending explicit paths, 
represented as bit-vectors, of candidate of nodes which get added to the sub-tree of the worker receiving 
these paths. Instead of representing paths as bit-vectors, our scheme represents them 
using \textit{test-depth pairs}. 
\emph{Simple Static Partitioning(SSP)\cite{jpfpar1}} applies parallel symbolic execution on Java bytecode. 
This technique builds on top of Symbolic PathFinder\cite{SPF} as the symbolic execution framework. 
SSP performs a shallow symbolic search and collects path constraints for the explored paths. 
These constraints are then distributed among workers to act as pre-conditions for symbolic execution. 
Disjoint and complete constraints ensure that the partitioning is non-overlapping and every path is visited. 
SSP requires processing the recorded constraints to collect the frequency of each variable 
and identifying them as \emph{cheap} or \emph{expensive}. Preference is given to constraints that 
are cheaper to solve. As the name suggests, this scheme does not support load balancing.

\section{Limitations and Future Work}
A significant challenge of using a distributed scheme like ours is that the outcomes 
are dependent on program characteristics.  As seen with the test subjects, most programs 
benefit from  partitioning and distribution while a few of them do not see significant improvements in 
execution speeds. We try to address this issue by making the test suite as diverse as possible, 
and we plan to study and gain insights as to the program features that determine its symbolic execution
performance 
in a distributing setting like ours. The current implementation of our scheme doesn't 
differentiate between execution paths when offloading work. As has been discussed in previous 
sections, the impact of distribution and load balancing depends on the size of the execution space and the 
nature of solver queries that lie in the region covered by a test-depth pair. To increase the 
efficacy of a scheme like test-depth partitioning, it is essential that only the paths, 
for which the communication overheads are more than compensated by added parallelism, be 
used for distribution. We attempt to alleviate this bottleneck by prioritizing shorter 
execution paths for transfers that happen due to load balancing. However, using program analysis 
techniques to devise intelligent partitioning methods is a research avenue we look forward to 
exploring. Many applications display different behaviour during different phases of execution.
Several subject programs used for experimentation comprise an initial phase where the inputs 
are parsed and checked for validity and then processed in subsequent stages. Another extension 
of the current work can be to identify phases of an application that benefit the most from parallelism 
and selectively distribute only those parts.

\section{Conclusions}
This paper presents a new approach to ranging based distributed symbolic 
execution called \emph{test-depth paritioning}. We build this system using 
a symbolic execution tool called \emph{KLEE}. The proposed scheme uses 
\emph{test-depth} pairs, representing partial paths, to define non-overlapping 
regions in the execution tree of a program. These pairs are given to the 
workers to carry out symbolic execution in parallel. Our system allows for 
load balancing using work-stealing in which partially explored paths are 
converted to test-depth pairs that are transferred from a busy worker to an 
idle worker. Our scheme overcomes a major limitation of previous work on ranging 
by removing the restriction of using only depth-first search to carry out 
symbolic exploration. We evaluate our implementation using twenty-five programs 
from GNU Coreutils and three application libraries - bc, osip, and libYAML. 
Our findings demonstrate \emph{test-depth partitioning} to be a promising 
distributed symbolic execution framework achieving median linear speed-ups when 
using depth-first and random-state searches with two workers and almost 
linear speed-ups of \emph{3.7x} with four workers for the Coreutils programs. 
The three libraries, using random-state search, witness median speed-ups of 
\emph{1.73x} and \emph{3.85x} with two and four workers respectively while a 
depth-first search provides median speed-ups of \emph{1.7x} and \emph{3.14x} 
for the two configurations respectively.

\bibliographystyle{splncs04}
\bibliography{references}
\end{document}